\documentclass[prl,aps,twocolumn,superscriptaddress,showpacs,preprintnumbers,amsmath,amssymb]{revtex4}
\usepackage{color}
\usepackage{graphicx}% Include figure files
\usepackage{ulem}
\usepackage{hyperref}
\def\textbf#1{\boldsymbol{#1}}

\begin{document}

\title{Dispersive spin excitations in highly overdoped cuprates revealed by resonant inelastic x-ray scattering}

\author{M.~Le Tacon}
\affiliation{Max-Planck-Institut~f\"{u}r~Festk\"{o}rperforschung,
Heisenbergstr.~1, D-70569 Stuttgart, Germany}

\author{M. Minola}
\affiliation{CNR-SPIN, Dipartimento di Fisica, Politecnico di Milano, I-20133 Milano, Italy}

\author{D. C. Peets}
\affiliation{Max-Planck-Institut~f\"{u}r~Festk\"{o}rperforschung,
Heisenbergstr.~1, D-70569 Stuttgart, Germany}

\author{M. Moretti Sala}
\affiliation{European Synchrotron Radiation Facility, BP 220, F-38043 Grenoble Cedex, France}

%\author{L. Hozoi}
%\affiliation{Institute for Theoretical Solid State Physics, IFW Dresden, Helmholtzstr. 20, 01069 Dresden, Germany}

%\author{Jeroen van den Brink}
%\affiliation{Institute for Theoretical Solid State Physics, IFW Dresden, Helmholtzstr. 20, 01069 Dresden, Germany}

\author{S. Blanco-Canosa}
\affiliation{Max-Planck-Institut~f\"{u}r~Festk\"{o}rperforschung,
Heisenbergstr.~1, D-70569 Stuttgart, Germany}

\author{V. Hinkov}
\affiliation{Max Planck-UBC Center for Quantum Materials, Vancouver, Canada}
\affiliation{Max-Planck-Institut~f\"{u}r~Festk\"{o}rperforschung,
Heisenbergstr.~1, D-70569 Stuttgart, Germany}

\author{R. Liang}
\affiliation{Department of Physics \& Astronomy, University of British Columbia, Vancouver, Canada}
\affiliation{Canadian Institute for Advanced Research, Toronto, Canada}

\author{D. A. Bonn}
\affiliation{Department of Physics \& Astronomy, University of British Columbia, Vancouver, Canada}
\affiliation{Canadian Institute for Advanced Research, Toronto, Canada}

\author{W. N. Hardy}
\affiliation{Department of Physics \& Astronomy, University of British Columbia, Vancouver, Canada}
\affiliation{Canadian Institute for Advanced Research, Toronto, Canada}

\author{C. T. Lin}
\affiliation{Max-Planck-Institut~f\"{u}r~Festk\"{o}rperforschung, Heisenbergstr.~1, D-70569 Stuttgart, Germany}

\author{T. Schmitt}
\affiliation{Swiss Light Source, Paul Scherrer Institut, CH-5232 Villigen PSI, Switzerland}

\author{L. Braicovich}
\affiliation{CNR-SPIN, Dipartimento di Fisica, Politecnico di Milano, I-20133 Milano, Italy}

\author{G. Ghiringhelli}
\affiliation{CNR-SPIN, Dipartimento di Fisica, Politecnico di Milano, I-20133 Milano, Italy}

\author{B.~Keimer}
\affiliation{Max-Planck-Institut~f\"{u}r~Festk\"{o}rperforschung,
Heisenbergstr.~1, D-70569 Stuttgart, Germany}

\date{\today}

\begin{abstract}
Using resonant inelastic x-ray scattering (RIXS) at the Cu $L$-absorption edge, we have observed intense, dispersive spin excitations in highly overdoped Tl$_2$Ba$_2$CuO$_{6+\delta}$ (superconducting $T_c =6$ K), a compound whose normal-state charge transport and thermodynamic properties have been studied extensively and shown to exhibit canonical Fermi-liquid behavior. Complementary RIXS experiments on slightly overdoped Tl$_2$Ba$_2$CuO$_{6+\delta}$ ($T_c =89$ K) and on Y$_{1-x}$Ca$_{x}$Ba$_2$Cu$_3$O$_{6+\delta}$ compounds spanning a wide range of doping levels indicate that these excitations exhibit energies and energy-integrated spectral weights closely similar to those of antiferromagnetic magnons in undoped cuprates. The surprising coexistence of Fermi-liquid-like charge excitations and high-energy spin excitations reminiscent of antiferromagnetic insulators in highly overdoped compounds poses a challenge to current theoretical models of the cuprates.
\end{abstract}

\maketitle

High temperature superconductivity arises when the CuO$_2$ planes of layered copper-oxide compounds are doped with mobile charge carriers. When the number of mobile carriers per Cu atom, $p$, vanishes, the CuO$_2$ planes are antiferromagnetically ordered and exhibit conventional spin wave excitations with a total bandwidth of $\sim 300$ meV. For hole doping $p \gtrsim 0.05$, the antiferromagnetic long-range order disappears, and the low-temperature ground state becomes superconducting. Inelastic neutron scattering (INS) experiments have demonstrated that dispersive spin excitations akin to antiferromagnetic spin waves persist in the superconducting state, although their low-energy spectral weight is progressively reduced with increasing $p$. According to current theories, these ``paramagnon'' excitations act as a key driving force for Cooper pairing \cite{Scalapino_RMP2012}. Up to now, however, the INS studies have been largely limited to underdoped ($0.05 \lesssim  p \lesssim 0.15$) \cite{Sidis_PSS2004,Birgeneau_JPSJ2006,Fujita_JPSJ2012} and lightly overdoped ($0.15 \lesssim p \lesssim 0.2$) cuprates, \cite{He_PRL2001,Pailhes_PRL2003,Capogna_PRB2007} where the spectral features attributable to paramagnons remain relatively sharp and intense. In the highly overdoped regime ($p > 0.2$), where the superconducting transition temperature, $T_c$, and the superconducting energy gap are sharply reduced and eventually vanish, ~\cite{Norman_Nature98,Miyakawa_PRL98,Le Tacon_NaturePhysics06} INS experiments have thus far only been reported for a single cuprate family, La$_{2-x}$Sr$_{x}$CuO$_4$ (La-214). \cite{Wakimoto_PRL2007,Lipscombe_PRL2007} They show progressive weakening of the spin excitations with energies $E \lesssim 100$ meV, continuing the trend already identified in the underdoped regime, but also indicate that excitations at higher $E$ are less affected by doping. Since La-214 exhibits incommensurate magnetic order (``stripes'') near optimal doping, and its $T_c$ is limited to $\sim 40$ K, it is unclear whether these findings are generic for the cuprate superconductors. Recent resonant inelastic x-ray scattering (RIXS) experiments \cite{letacon_NatPhys2011,Dean_condmat} allowed the detection of dispersive high-energy ($E \gtrsim 100$ meV) spin excitations in optimally doped cuprates with maximal $T_c \sim 90$ K, as well as in iron-based superconductors~\cite{Zhou_NatCom2013}. Although these high-energy paramagnons are strongly broadened by scattering from mobile carriers and thus difficult to detect by INS, their energies and integrated spectral weights turned out to be surprisingly similar to those of magnons in antiferromagnetically ordered compounds with $p=0$. We have now used the methodology established in these experiments to explore the highly overdoped regime of the cuprate phase diagram.
%, where bi-magnon excitations have recently been reported~\cite{Bisogni_PRB2012,Li_PRL2012}.

Most of our current knowledge about highly overdoped cuprates derives from transport and thermodynamic experiments, which have revealed characteristics matching those of ordinary Fermi liquids. In particular, the electrical resistivity essentially depends quadratically on temperature~\cite{Kubo_PRB91, MacKenzie_PRB1996}, the thermal conductivity satisfies the Wiedemann-Franz law~\cite{Proust_PRB05}, and the uniform magnetic susceptibility is dominated by a temperature independent, Pauli-like term~\cite{Bobroff_hab}. In highly overdoped Tl$_2$Ba$_2$CuO$_{6+\delta}$ (Tl-2201), a compound that features a single, isolated CuO$_2$ plane per formula unit and very low intrinsic disorder, additional photoemission ~\cite{Plate_PRL05}, angle-dependent magnetoresistance \cite{Hussey_Nature03}, and quantum oscillation~\cite{Vignolle_Nature08} experiments have uncovered well-defined Landau quasiparticles with a Fermi surface that agrees quantitatively with the predictions of density functional theory. Based on these findings, one commonly assumes that the spin excitations in highly overdoped cuprates resemble those of weakly correlated metals, which exhibit a featureless continuum of incoherent electronic spin-flip excitations extending up to the Fermi energy, rather than the dispersive paramagnon excitations found in the cuprates at lower doping levels.

We report a RIXS study of the model system Tl-2201 that challenges this commonly held view of the highly overdoped cuprates. Specifically, RIXS data on a Tl-2201 crystal with $T_c = 6$ K (corresponding to hole content $p = 0.27$ according to Tallon's empirical formula~\cite{Tallon_PRB1995}), which is squarely in the ``Fermi liquid'' regime of the phase diagram, reveal intense spin excitations with a dispersion relation
closely similar to antiferromagnetic spin waves.
%Complementary experiments on optimally doped Tl-2201 ($p = 0.17$, $T_c = 90$ K), and on (Y,Ca)Ba$_2$Cu$_3$O$_{6+\delta}$ (Y-123) crystals spanning %the doping range $0 \leq p \leq 0.19$, demonstrate that the energies and energy-integrated spectral weights of the highest-energy paramagnons %remain
This implies that short-range spin correlations are surprisingly robust even in cuprates with well-documented, canonical Fermi-liquid behavior at low energies.

\begin{figure}
\includegraphics[width=\linewidth]{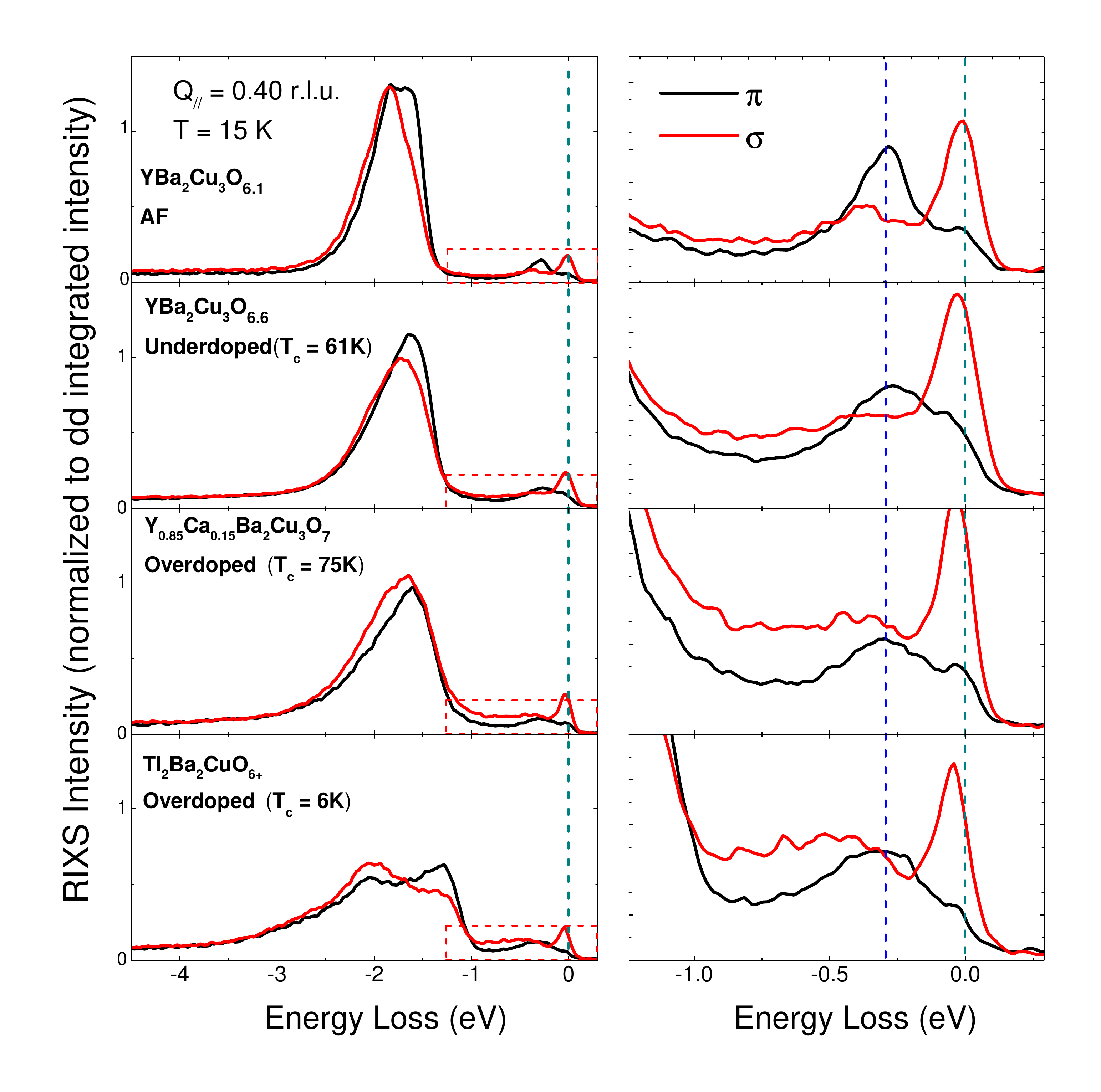}
\caption{(Color online) Left column: Doping dependence of the RIXS response measured in the $\sigma$ (red) and $\pi$ (black) scattering geometries for ${\bf Q_{//}} = 0.8 \Gamma X =(0.4, 0)$ r.l.u. Right column: Detailed view of the low energy part of the data. In each panel, the intensity scale has been normalized to the area of the $dd$ excitation in the $\pi$ channel.}
\label{Fig1}
\end{figure}

The RIXS measurements were performed at the ADRESS beamline~\cite{Strocov_JSR2010} of the Swiss Light Source (Paul Scherrer Institute,
Switzerland) using the high-resolution SAXES spectrometer~\cite{Ghiringhelli_RSI2006}. To obtain the resonant conditions required for the observation of magnons, the energy of the incident x-rays was tuned to the maximum of the $L_3$ absorption peak of the planar Cu atoms ($\sim$ 931.5 eV). Since the absorption edge of Cu atoms on chain sites in the Y-123 structure is at a different energy, these atoms do not contribute to the spectra reported there. The total energy resolution was about 130 meV, and the position of the elastic (zero energy loss) line was determined by measuring a non-resonant spectrum of polycrystalline graphite for each value of the momentum transfer, $Q$. The spectra presented here were recorded at $T=15$ K for total durations of 30-120 minutes.
%, as the sum of individual spectra of 5 min duration.
Scattering of photons at the Cu $L_3$ edge is restricted to a maximum $Q = 0.855$ \AA$^{-1}$, which covers about 85$\%$ of the first Brillouin zone (BZ) along the [100] direction. Momentum transfers are given in reciprocal lattice units (r.l.u.), that is, in units of the reciprocal lattice vectors $a^*$=2$\pi/a$, $b^*$=2$\pi/b$, and $c^*$=2$\pi/c$, where $a$, $b$, $c$ are the dimensions of the tetragonal (orthorhombic) units cells of Tl-2201 (Y-123). Because of the quasi-two-dimensional electronic structure, we mostly refer to the projection of $Q$ parallel to the CuO$_2$ layers, $Q_{//}$.

\begin{figure*}
\includegraphics[width=0.95\linewidth]{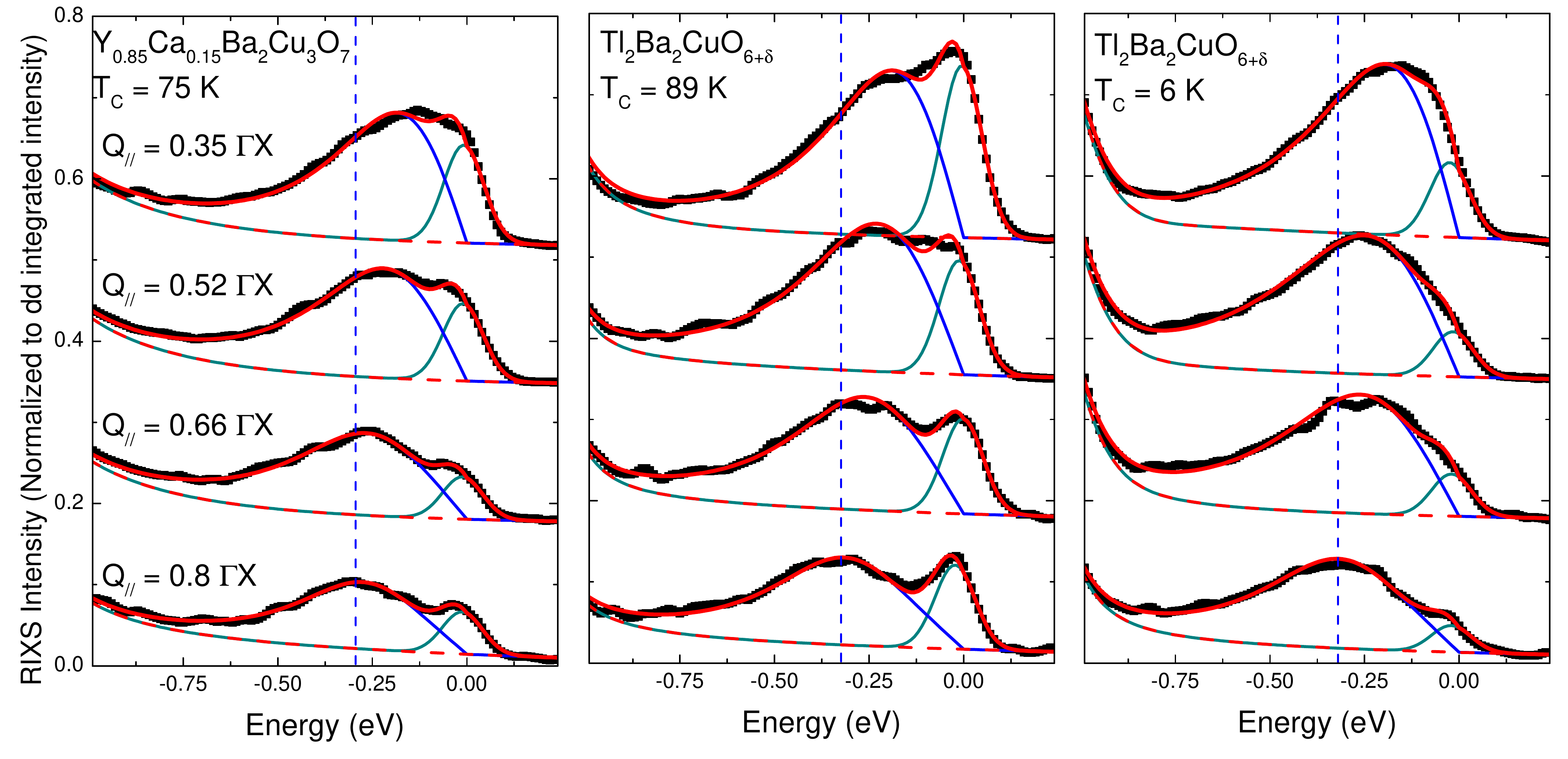}
\caption{(Color online) Low energy part of the RIXS spectra of overdoped Y$_{0.85}$Ca$_{0.15}$Ba$_2$Cu$_3$O$_{6+x}$ (left panel), moderately overdoped Tl$_2$Ba$_2$CuO$_{6+\delta}$ ($T_c$ = 89 K, middle panel) and strongly overdoped Tl$_2$Ba$_2$CuO$_{6+\delta}$ ($T_c$ = 6 K) in the $\pi$ scattering geometry for various in-plane momentum transfers $Q_{//}$. The fitting procedure is detailed in Ref.~\cite{letacon_NatPhys2011}. The slightly more intense elastic line in the $T_c = 89$ K sample, compared to the other two samples, originates from the lower quality of its surface. The data were normalized to the energy-integrated spectral weight of of the $dd$ excitations, which is proportional to the density of CuO$_2$ layers.}
\label{Fig2}
\end{figure*}

Tl-2201 single crystals of typical size $0.5 \times 0.5 \times 0.03$ mm$^3$ were grown by an encapsulated copper-rich self-flux technique and annealed under controlled oxygen partial pressures to yield the desired doping levels.~\cite{Peets2010} The experiments were performed on a lightly ($p$ = 0.17, $T_c=89$ K) and a heavily ($p$ = 0.27, $T_c =6$ K) overdoped crystal. Y-123 crystals of typical size $2 \times 2 \times 0.1$ mm$^3$ were grown by a top-seeded solution growth method \cite{Lin_JCG2002}. We report new RIXS data on an overdoped Y$_{0.85}$Ca$_{0.15}$Ba$_2$Cu$_3$O$_{6+x}$ single crystal ($p= 0.19$, $T_c = 75$ K), and on an antiferromagnetically ordered YBa$_2$Cu$_3$O$_{6.1}$ crystal ($p \sim 0$). For comparison, we also show data on a YBa$_2$Cu$_3$O$_{6.6}$ sample reported previously. \cite{letacon_NatPhys2011}

Figure~\ref{Fig1} shows representative RIXS spectra of Y-123 and Tl-2201 crystals spanning a wide range of doping levels. The spectra were taken for $Q_{//} = (0.4, 0)$, close to the $X$ point, $(0.5, 0)$, at the BZ boundary, with the polarization of the incident photon either in the scattering plane ($\pi$ scattering geometry) or perpendicular to it ($\sigma$ geometry). In the energy range $\sim 1-3$ eV, we observe intense interband transitions (Fig.~\ref{Fig1}, left panel), which are known as $dd$-transitions because the initial and final states are of Cu $d$-orbital character. The shapes of the $dd$ excitation spectra and their polarization dependence reflect the local environment of Cu, and are hence quite different in the Tl-2201 and Y-123 families.~\cite{Braicovich_PRL2010,letacon_NatPhys2011,Moretti_NJP2011} A detailed assignment of these features can be made with the help of quantum chemistry calculations \cite{Hozoi_ScientificReports2012}, which will be reported elsewhere.

Here we focus on the low-energy features in the spectra (right panel of Fig.~\ref{Fig1}), which are independent of the chemistry and lattice symmetry of the two different compound families. The spectrum of YBa$_2$Cu$_3$O$_{6.1}$ shows a resolution-limited single-magnon peak whose energy ($E \sim 260$ meV) matches the one previously determined by INS and RIXS experiments on the 123 family. \cite{Hayden,letacon_NatPhys2011} Since these data were obtained close to the $X$ point, single spin-flip excitations are only visible in the $\pi$ geometry.~\cite{Ament_PRL2009} In the $\sigma$ channel, a weaker feature due to multiple-magnon excitations is present at higher energy. When mobile holes are added to the system, a continuum of electron-hole pair excitations appears in the $\sigma$ channel. Its intensity continuously increases with increasing doping. In the $\pi$ channel, on the other hand, the single-magnon peak remains centered at the same position, although it broadens considerably due to interactions with the electron-hole excitations. The key observation is that the peak associated with these excitations remains well-defined and centered at the same energy even in a highly overdoped Tl-2201 specimen with $p = 0.27$. In fact, the $\pi$-channel spectra of the overdoped Y-123 and Tl-2201 samples look remarkably similar to the one of the underdoped YBa$_2$Cu$_3$O$_{6.6}$ sample also shown in Fig.~\ref{Fig1}.

\begin{figure}
\includegraphics[width=\linewidth]{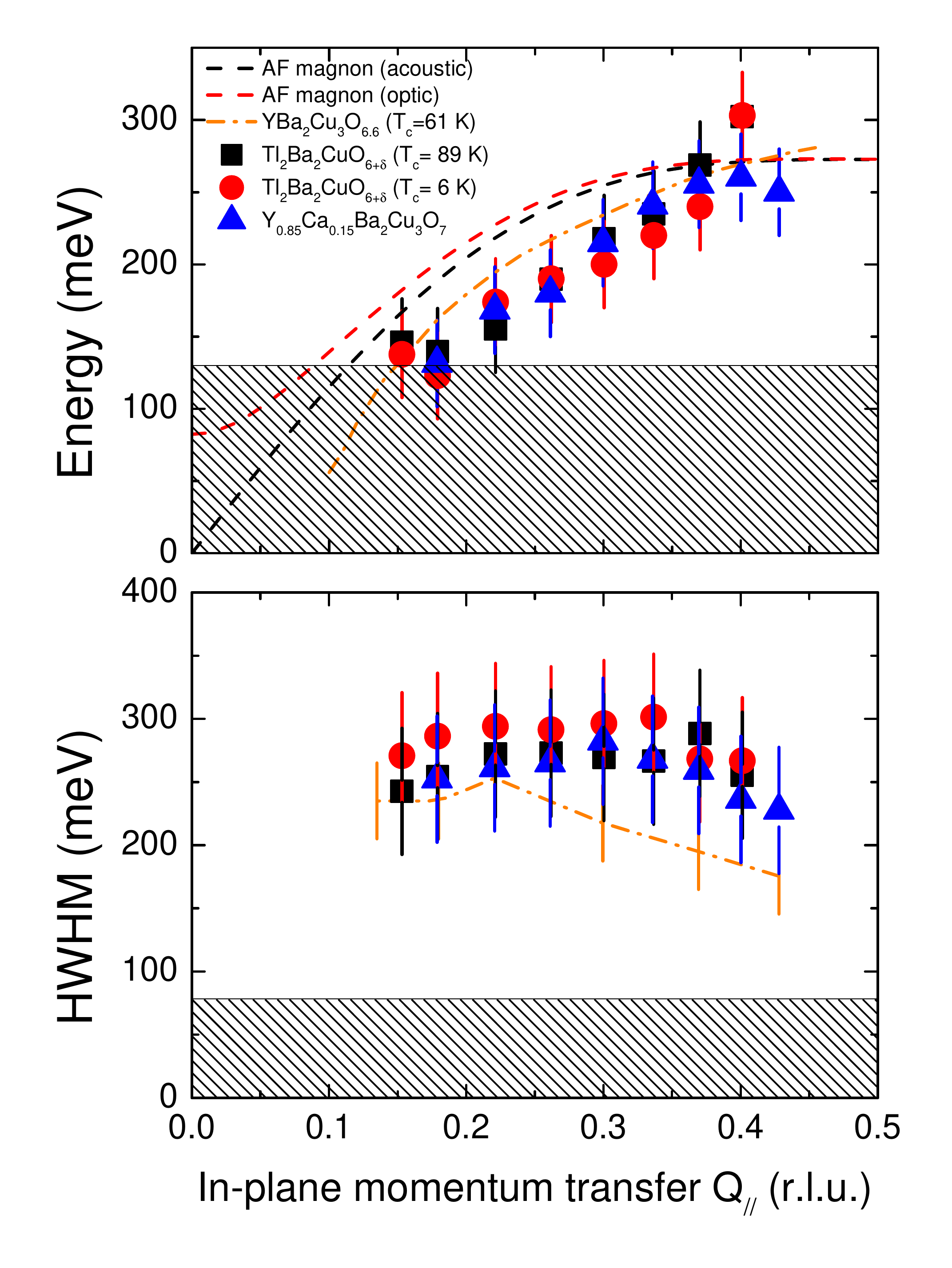}
\caption{(Color online) Energy and  half-width-at-half-maximum (HWHM) of the magnetic excitations in the three overdoped compounds as function of the in-plane momentum transfer $Q_{//}$. The dashed lines represent the dispersion of acoustic (black) and optic (red) magnons in the antiferromagnetically ordered compound, whereas the dashed-dotted line indicates the energy and HWHM of underdoped YBa$_2$Cu$_3$O$_{6.6}$ taken from Ref.~\cite{letacon_NatPhys2011}. In both panels, the dashed area corresponds to the experimental energy resolution. The data were normalized to the energy-integrated spectral weight of the $dd$ excitations.}
\label{Fig3}
\end{figure}

We next address the momentum dependence of the paramagnon excitations. Figure~\ref{Fig2} displays RIXS spectra of the three overdoped samples in the $\pi$ channel for various momentum transfers along the $\Gamma-X$ direction. The systematic shift of the peak center away from 260 meV (dashed line in Fig.~\ref{Fig2}) clearly demonstrates the dispersive nature of these modes. This trend is confirmed by fits of the data to a set of Lorentzian profiles, following Ref.~\onlinecite{letacon_NatPhys2011}. The results yield excellent descriptions of the data (blue lines in Fig.~\ref{Fig2}).

Figure~\ref{Fig3} shows the energies and linewidths of the paramagnon features in the three overdoped compounds extracted from the Lorentzian fits. The dispersion relations of the magnetic excitation in the overdoped samples resemble those of antiferromagnetic magnons in undoped cuprates, and the shapes of the paramagnon peaks are very similar to those in the samples at lower doping levels ~\cite{Braicovich_PRL2010,letacon_NatPhys2011}, although a slight softening and broadening is apparent relative to the underdoped YBa$_2$Cu$_3$O$_{6.6}$ sample (shown for comparison as an orange line in Fig. ~\ref{Fig3}). Interestingly, we do not observe sizeable differences between the data on the lightly and strongly overdoped Tl$_2$Ba$_2$CuO$_{6+\delta}$.

%For the momentum transfers ${\bf Q_{//}} > 0.25$, the integrated intensity of the low-energy excitation is essentially doping independent, while, for ${\bf Q_{//}} < 0.25$  it increases with doping. This is caused by an increase of the cross section for the electron-hole excitations in the $\pi$ channel at low q~\cite{Ament_PRL2009}, and only an analysis of the scattered light polarization, not technically possible to date, could in principle allow the separation from spin and charge excitations.
%The results are reported on the right panel of Fig.~\ref{Fig3}: dispersion and linewidth in the two samples are essentially identical, and very similar to  those of the paramagnon excitation observed in YBCO. This shows that these excitations remain present up to the highest doping levels. Even more surprising, the integrated intensities of the excitation are the same for the two doping levels, and overall a factor of 2 weaker than in YBCO, which is consistent with the fact that the Tl2201 compounds have only one single CuO$_2$ layer per unit cell.

\begin{figure}
\includegraphics[width=\linewidth]{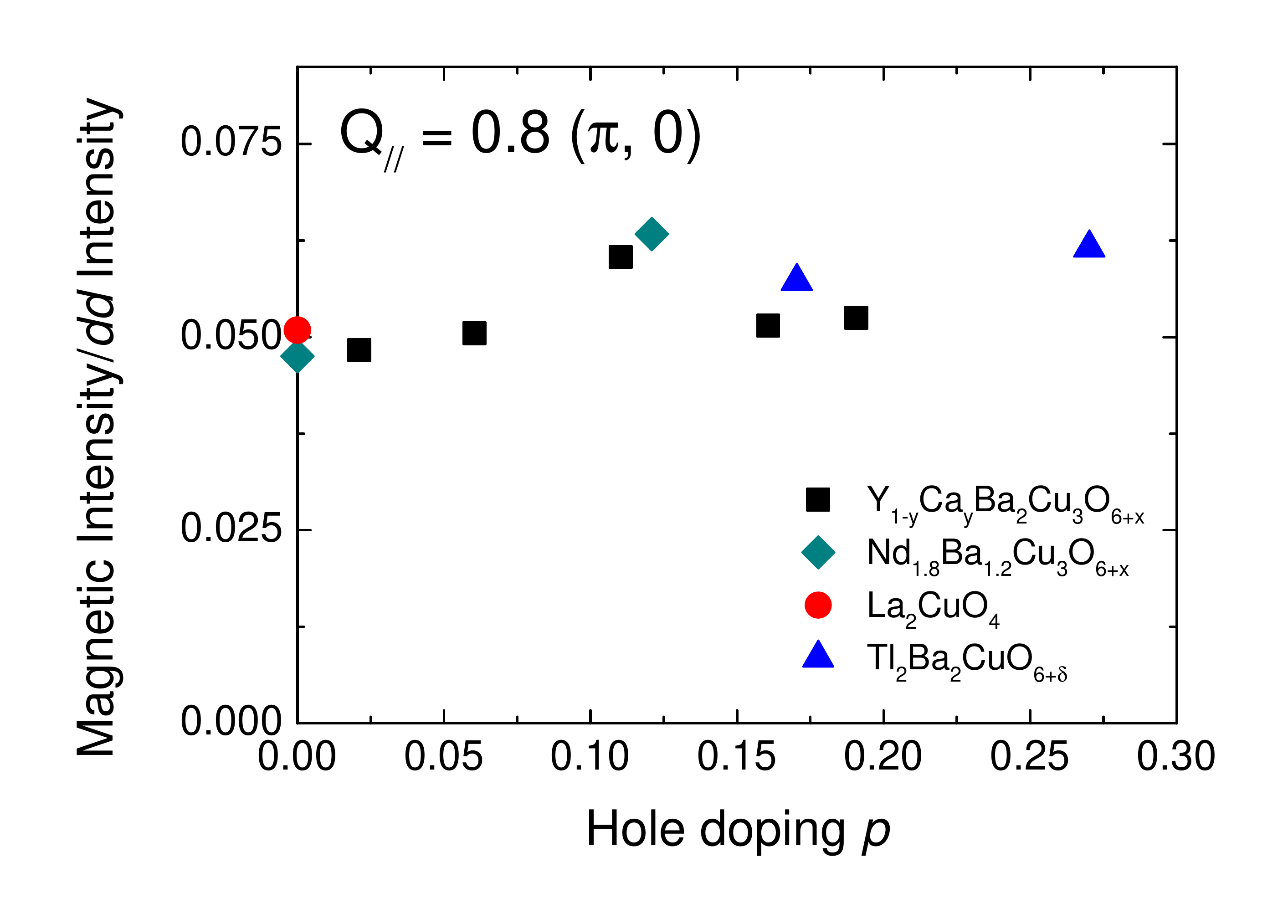}
\caption{(Color online) Energy-integrated intensity of the magnetic excitations, normalized to the $dd$ excitation intensity, as a function of doping.}
\label{Fig4}
\end{figure}

In order to quantitatively compare the doping dependence of the magnetic spectral weights in the Tl-2201 and Y-123 systems with one and two CuO$_2$ layers per formula unit, respectively, we have normalized the RIXS intensity to the energy-integrated spectral weight of the $dd$-excitations, which is proportional to the density of CuO$_2$ layers. Inspection of Figs.~\ref{Fig2} and~\ref{Fig3}, where the intensity scale has been calibrated in this way, shows that the paramagnon spectral weight per Cu atom is closely similar in both compounds. In Fig.~\ref{Fig4} we plot the normalized, energy-integrated paramagnon intensity close to the BZ boundary, where most of the intensity in the $\pi$ channel is of magnetic origin, as a function of doping for both Tl-2201 and Y-123. We have also included data for antiferromagnetically ordered La$_2$CuO$_4$ taken from Ref.~\cite{Braicovich_PRL2010}. Clearly, our previous conclusion~\cite{letacon_NatPhys2011} that the high-energy magnetic excitations in the cuprates are doping independent continues to hold deep into the overdoped regime.

In summary, our data demonstrate that high-energy spin excitations with dispersion relations and energy-integrated spectral weights closely similar to antiferromagnetic magnons persist up to doping levels sufficient to almost entirely suppress superconductivity. They corroborate and extend prior INS~\cite{Lipscombe_PRL2007} and oxygen $K$-edge RIXS~\cite{Bisogni_PRB2012} results on La$_{1.78}$Sr$_{0.22}$CuO$_4$ ($p = 0.22$), which revealed magnetic excitations up to $E \sim 160$ meV and bimagnon modes up to $E \sim 450$ meV, respectively. The behavior we have observed is reminiscent of the persistence of high-energy magnons upon heating well into the paramagnetic regime of magnetically ordered insulators such as EuO,~\cite{Boni} which is well understood as a consequence of short-range correlations between local moments. This analogy suggests an interpretation of our cuprate data in terms of short-range correlations between Cu moments. The main result of our study is that these correlations remain surprisingly strong even in highly overdoped Tl-2201, a compound whose charge transport and thermodynamic properties are well described by Fermi-liquid theory. \cite{Plate_PRL05,Hussey_Nature03,Vignolle_Nature08} The development of a unified framework that consistently describes both the fermiology and the spin dynamics of this prototypical overdoped cuprate is an important challenge to theory.

We end our discussion with some remarks on the implications of our data for spin-fluctuation mediated Cooper pairing theories. Recent optical spectroscopy experiments~\cite{vanHeumen_PRB2010,DalConte_Science2012} indicate that the conduction electrons in the cuprates are coupled to bosonic excitations with a spectrum extending up to energies comparable to zone-boundary magnons in insulating cuprates, but they suggested that coupling to high-energy excitations is present even in highly overdoped, non-superconducting cuprates where magnon-like excitations were not expected.~\cite{vanHeumen_PRB2010} Our observation of high-energy paramagnons in the highly overdoped regime now resolves this puzzle. The persistence of  substantial electron-paramagnon coupling in overdoped cuprates with greatly depressed superconductivity is qualitatively consistent with phenomenological models according to which the highest-energy spin excitations are pair-breaking in the $d$-wave channel. \cite{Scalapino_RMP2012,letacon_NatPhys2011,Onufrieva_PRL2012} In these models, the $d$-wave pairing strength comes from lower-energy spin excitations centered around the antiferromagnetic ordering wave vector $Q_{//} = (0.5, 0.5)$. Close to this wavevector, low-energy ($E \sim 40-60$ meV) spin excitations have indeed been observed by INS in many optimally doped cuprates \cite{Sidis_PSS2004,Birgeneau_JPSJ2006,Fujita_JPSJ2012} including Tl-2201.~\cite{He_Science} INS data on La-214 \cite{Wakimoto_PRL2007,Lipscombe_PRL2007} indicate that these low-energy excitations build up progressively as the optimal doping level is approached from above, in concert with the Cooper pairing correlations. Together with prior studies, our RIXS data are thus beginning to provide a firm experimental basis for a controlled theoretical approach to the high-$T_c$ problem from the overdoped side, where difficulties due to competing spin~\cite{Haug_NJP2010} and charge~\cite{Ghiringhelli_Science2012} order are much less severe than in the underdoped regime on which much of the experimental effort has focused in the past.

\noindent \textit{Note added}. Intense magnetic excitation have recently also been observed by M.P.M. Dean and collaborators in highly doped La-214 ~\cite{Dean_private}.

\noindent \textit{Acknowledgement}. We thank J. Chaloupka, A. Damascelli, J.P. Hill, L. Hozoi, G. Khaliullin, F. Onufrieva, P. Pfeuty, G. A. Sawatzky, O. Sushkov, and J. van den Brink for valuable discussions, and K. Zhou and C. Monney for technical support. This work was partly supported by the European project SOPRANO (Grant No. PITN-GA-2008-214040).

\end{document}